\documentstyle[epsf,12pt,a4]{article}
\newcommand{\be}{\begin{equation}}
\newcommand{\ba}{\begin{eqnarray}}
\newcommand{\ee}{\end{equation}}
\newcommand{\ea}{\end{eqnarray}}
\newcommand{\nn}{\nonumber}

\newcommand{\GeV}{\;\mbox{GeV}}
\newcommand{\MeV}{\;\mbox{MeV}}
\newcommand{\eV}{\;\mbox{eV}}
\newcommand{\keV}{\;\mbox{keV}}

\newcommand{\secs}{\;\mbox{s}}

\newcommand{\simgt}{\stackrel{>}{{}_\sim}}
\newcommand{\ol}{\overline}

\newcommand{\yr}{\;\mbox{yr}}
\newcommand{\Gyr}{\;\mbox{Gyr}}

\newcounter{currequation}
\begin{document}
\title{
An asymptotic decrease of $(m_p/m_e)$ with cosmological time, from a decreasing, small effective vacuum
expectation value moving from a maximum in the early universe
}
\author{Saul Barshay\footnote{barshay@kreyerhoff.de} and Georg Kreyerhoff\footnote{georg@kreyerhoff.de}\\
III.~Physikalisches Institut A\\
RWTH Aachen\\
D-52056 Aachen}
\maketitle
\begin{abstract}
The empirical, possible small variation downward by about $10^{-5}$, of the ratio of the proton mass
to the electron mass, over a characteristic time interval estimated here to be about a billion years,
is related to the decrease with time of a small, effective vacuum expectation value for a Goldstone-like
pseudoscalar field which is present in the early universe.
The same vacuum expectation value controls the magnitude of a very small, residual vacuum energy density
today, which has decreased only a little in this model. The present time variation is estimated to be near to a definite
limit, that is asymptotically approached as $t\to \infty$.
\end{abstract}
Recently, new data \cite{ref1} has given an indication that the ratio of the proton to electron
mass $\mu = (m_p/m_e)$, has decreased over a cosmological time interval. If interpreted in terms of
an effective decrease in the proton mass, the data suggest a decrease by about $(10 \eV)\times \mu
\sim 18 \keV$, over a period of about the past twelve billion years. Natural questions which arise
are then the following.
\begin{itemize}
\item[(1)] Can one obtain the necessity for, and the direction of change, a decrease, independently
of the possibility that coupling parameters such as $\alpha_{\mathrm{em}}$, depend upon time? Data
on  the latter possibility \cite{ref2,ref3}, have stimulated the search for time variation of
physical ``constants'' \cite{ref4,ref1}. Recent data \cite{ref5,ref6} have not yet confirmed a
variation \cite{ref2,ref3}. Further data are forthcoming.
\item[(2)] Can one obtain an estimate of the small absolute scale of the mass decrease, a few
keV, by relating it to other small energy-scale effects, like a present, residual vacuum energy
density (effective cosmological constant) of about $3\times 10^{-47} \GeV^4$, and possibly to
a neutrino mass of less than 0.1 eV?
\item[(3)] Is it possible to show that the time variation is approaching an asymptotic limit,
as $t\to \infty$?
\item[(4)] Is it possible to estimate the cosmological time scale $\overline{t}$, for significant
mass decrease (on the above small energy scale), in terms of another cosmological parameter 
associated with the early universe?
Possibly a significant decrease has also occurred in the first two billion years of the
universe, as well as in the subsequent twelve billion years (present
universe age $\sim 14 \Gyr$). 
\end{itemize}
The purpose of this note is to remark that there
might be affirmative answers to all of these questions, and to make simple, but 
quantitative estimates of the relevant numbers. The basis for the results is the standard
assumption that the main contributions to particle mass arise from the non-zero
vacuum expectation values of spin-zero fields. It is usually assumed that such a
vacuum expectation value arises at the stable minimum of some effective potential-energy
density. Here, the first assumption leading to the results below, is that a small, particular
vacuum expectation value of a pseudoscalar, Goldstone-like field $b$ (which spontaneously
breaks CP invariance in a cosmological context) arises at a metastable (i.~e.~over
a cosmological time interval $\overline{t}$) maximum of an effective potential-energy
density \cite{ref7}. We note that the possibility that the standard, scalar-field 
inflationary dynamics in the very
early universe originates at a maximum of an effective potential for a (classical)
scalar (inflaton) field $\phi$, has already been considered in detail \cite{ref8,ref9,ref10}.
It has been shown that radiative corrections to a $\lambda \phi^4$ potential-energy density
can set up an effective energy density with a maximum at or just above the Planck mass
$M_{\mathrm{Pl}} \cong 1.2 \times 10^{19}\GeV$, {\underline{and a minimum}} just below
$M_{\mathrm{Pl}}$ \cite{ref9}. Inflation occurs 
with the inflaton field at the maximum and during its movement
to the minimum. The positive second derivative of the effective potential with respect
to the field variable, at the minimum, corresponds to the large squared mass of inflaton
quanta (estimated \cite{ref7,ref8} to be $m_\phi^2 \sim (5\times 10^{11}\GeV)^2$). If metastable
(i.~e.~essentially decoupled from big-bang radiation) these quanta can constitute dark
matter today \cite{ref7,ref8,ref11}. The pseudoscalar field $b$ can be connected with the
scalar $\phi$, in a hypothetical, idealized model of a cosmological, spontaneously-broken chiral
symmetry \cite{ref7,ref8}. However, the pseudoscalar field $b$ is a separate hypothesis from the
scalar field $\phi$ whose vacuum energy density generates the hypothetical inflation near $t=0$,
and the $b$-field dynamics over cosmological time intervals is distinct.
The hypothetical, small vacuum expectation value $F_b$ of the $b$ field,
contributes a residual vacuum energy density of magnitude $|\rho_\Lambda| = |-\lambda F_b^4|
\sim 2.7 \times 10^{-47} \GeV^4$ for $F_b\sim 5.5\eV$ \cite{ref7}, using the same, empirical \cite{ref12,ref13}
value of the self-coupling parameter as for the $\phi$ field, $\lambda\sim 3\times 10^{-14}$. An attempt is made to obtain
a separate estimate of $F_b$ by coupling $b$ to $\nu_\tau$ (with $g_{\nu_\tau}$) \cite{ref7}.
This gives rise to a (presumably largest) neutrino mass $m_{\nu_\tau} = \sqrt{\tilde{m}_{\nu_\tau}^2 +
(g_{\nu_\tau}F_b)^2} \sim \sqrt{(g_{\nu_\tau} F_b)^2} \sim 0.055\eV$, for $F_b \sim 5.5 \eV$ and
$g_{\nu_\tau} \sim 10^{-2}$ \cite{ref7}, and ``bare'' neutrino mass $\tilde{m}_{\nu_\tau} \sim 0$.
This provides a quantitative representation (including the significant role of $\lambda$) of the
often-remarked similarity between the empirical energy scales relevant for neutrino mass
and for an effective cosmological constant. The above negative sign of the vacuum energy
density can be changed to positive, by considering an (explicitly symmetry-breaking) effective
energy density for the $b$ field to be $(1/2 (\mu_b^2 b^2) - \lambda b^4)$ with $\mu_b^2>0$, instead
of  $(1/2 (\mu_b^2 b^2) + \lambda b^4)$ with $\mu_b^2 < 0$. There is then a maximum \cite{ref7} at
$F_b$, with $F_b^2 = \mu_b^2 /4\lambda$. The second derivative, the effective squared mass
is $(\mu_b^2 - 12\lambda F_b^2) = -8 \lambda F_b^2 < 0$. Thus, the second assumption here,
which allows for the following results, is that quanta with negative squared mass (i.~e.~superluminal
tachyons \cite{ref14,ref15}) are not present. This effectively prohibits strong long-range forces
due to  exchange of $b$ quanta \footnote{
There can be exchanged, pseudoscalar $b$ quanta for a brief time near to
$\sim 10^{-36}\secs$, if the $b$ field moves from zero to $F_b$. These can give a
CP-violating effect such as an antineutrino-neutrino asymmetry from a primary,
radiation-producing decay process \cite{ref7,ref11}. Spontaneous $CP$ violation is a 
motivation for considering a nonzero vacuum expectation value for the $b$ field.
}. Thus, we assume that the main effect of a hypothetical coupling to quarks of a 
classical $b$ field is to give a mass contribution to primordial quarks.
A coupling $g$ of the $b$ field to primordial ordinary quarks
gives a quark mass contribution of $gF_b$; 
subsequently for three confined valence quarks, a nucleon mass contribution of $3gF_b$.
We have assumed that primordial quarks have zero bare mass, and we do not consider
possible thermal effects in this paper. We assume that it is at a later time $\sim 10^{-12}\secs$, that
electroweak symmetry-breaking generates the standard-model MeV mass contribution for light quarks,
and we assume that this contribution then simply adds to the mass contribution estimated here, $gF_b$. 
This is the assumption that the electroweak mass term arises from the Higgs vacuum expectation
value times a tiny coupling to the quark field which has acquired a mass term $gF_b$.
Due to 
quark-antiquark pairs in the nucleon, the factor of 3 for the nucleon mass contribution,
can probably be considered to be a minimal enhancement factor.

The above (standard) simple example of an effective potential-energy density, is used here
only to illustrate a maximum, at $F_b$, and a point of zero second derivative, at $F_b/\sqrt{3}$. A
classical field can remain at the maximum, i.~e.~with $db/dt=0$. If we consider a very small
perturbation which causes a variation with time of the effective $F_b$, then the direction
of change must be assumed to be downward, so as to keep the related vacuum energy density
bounded as $t\to \infty$. For the above example, the equation of motion implies that $b(t)=F_b(t)$
approaches zero as $t$ increases, where $F_b(t)$ is here used to denote an effective vacuum
expectation value in different cosmological epochs. We must assume that this does not
happen. Thus, the essential physical assumption is that $F_b(t)$ moves from a largest value
at $t\sim 0$, toward zero, over cosmological time intervals parameterized by $\ol{t}$,
but comes only a part of the way as $t\to\infty$, such that the second derivative then
approaches zero through negative values. Although here, we have not obtained this from
some hypothetical, effective potential, the idea is that such a behavior can be a physical
possibility, which warrants being noted. Here we illustrate this possibility.
We consider numerically a simple example for such a possible time dependence for the field
$b(t) = F_b(t)$. 
\be
F_b^2(t) = \epsilon F_b^2(0) + \frac{(1-\epsilon) F_b^2(0)}{\left(1+\left(t/\overline{t}\right)^2\right)}
\ee
with the parameter $\overline{t} \sim 2 \times 10^9 {\mathrm{yr}}$. 
The limiting value is $\sqrt{\epsilon} F_b(0)$ as $t\to \infty$; $0<\epsilon < 1$.
So, we have
\ba
{\mathrm{at}}\;\; t\simgt 10^{-36}\secs && 3gF_b(0) \sim g (28.6\eV)\nn\\
{\mathrm{at}}\;\; t\sim \overline{t} && 3g\sqrt{(1+\epsilon)/2} F_b(0) \sim g (23.4\eV)\\
{\mathrm{at}}\;\; t\simgt 14\times 10^9 \yr && \sim 3g \sqrt{\epsilon} F_b(0) \sim g (16.5\eV)\nn
\ea
Only for orientation, we have given the above numerical values using $\sqrt{\epsilon}\sim 1/\sqrt{3}$.
From $t\simgt 10^{-36} \secs$ to 2 Gyr, there is a decrease of $-g (5\eV)$ in $m_p$. From $\sim 2 \Gyr$ to
$\simgt 14 \Gyr$, there is a decrease of $-g (7\eV)$. The asymptotic limit is approached
in the present ``old'' universe. 
There are positive answers to the first three questions
in the introduction. The direction of mass change is downward in the model.
This direction is independent of the resolution of the issue of possible variation 
of coupling parameters with time. There is a limiting decrease as $t\to \infty$; this
is related to the absence of a long-range force. Even  with a sizable effective
``magnification'' factor \footnote{
With reference to possible ``magnification'' of the effective $g$, it might be useful
to note that the confinement of quarks at $\sim 10^{-6}\secs$, does involve electroweak
mass $\sim \mbox{MeV}$, being substantially increased to constituent quark mass. The QCD
energy-scale parameter is $\Lambda \sim 220 \MeV$.
}, $g\sim 10^3$, one obtains a small scale of
mass change, $\sim \mbox{keV}$. Conceptually, this is related to a very small, residual
vacuum energy density in the present epoch,
and possibly to a very small neutrino mass. 
The electron mass can change, but the leptonic $b$ coupling
may be like that estimated for neutrinos, $g_l \sim 10^{-2}$. Thus, the hypothetical
downward change in $(m_p/m_e)$ is probably controlled by the downward change in $m_p$.
The parameter $\overline{t}$ is possibly related to other dynamical quantities in the
early universe. It might be connected with the ratio of vacuum expectation values,
which ratio is numerically closely given in terms of 
the very small parameter $\lambda$ \cite{ref11},
that scales the primordial, vacuum energy densities: $\lambda^2 \sim F_b(0)/\sqrt{3}\phi_c
\sim 5.5\eV/10^{18}\GeV \sim 5.5\times 10^{-27}$, for an inflaton mass $m_\phi
\sim 2 \sqrt{2} \sqrt{\lambda} \phi_c \sim 7.7\times 10^{11}\GeV$. (Here $\phi_c \sim 10^{18}\GeV$,
at an assumed minimum of zero for the inflaton effective potential.) 
Using the magnitude of the square root of the second derivative of the effective potential, $|m_b|$, 
as an estimate for a minimal fluctuation $|\delta b|$ of a (classical) $b$ field at a potential
maximum, and representing $|\delta b|$ as a geometric mean of the inverse of ``initial'' and
``final'' times, 
suggests $|\delta b| \sim \sqrt{(1/10^{-36}\secs) \times (1/\overline{t})} >
2\sqrt{2}\sqrt{\lambda} F_b(0) = |m_b|$. Then, $\overline{t} < (10^{-36}\secs) \times
(1/\lambda^4) \sim 3\times 10^{16}\secs$, where $10^{-36}\secs \sim 1/m_\phi \sim
\lambda^2/|m_b|$, is the time near the end of inflation generated by the $\phi$ field \cite{ref11} \footnote{
As counted from the time of $\phi$ leaving its value $\simgt M_{\mathrm{Pl}}$ at the effective potential maximum.
There is
no time-reversal symmetry for the vacuum in this model because of the presence of the $b$ field (odd under time reversal), and
of a discontinuity in its behavior at $t\sim 0$.
}.
With this value for $\ol{t}$, the expansion scale factor given approximately as 
$a(\overline{t}) \sim (\overline{t}/10^{-36}\secs)^{1/2} \sim (1/\lambda^2)$ \footnote{It is interesting to note that the
necessary minimal value of the expansion scale factor for an initial inflation over
$\Delta t$ is a similar number. That is $a_{\mathrm{infl}} (\Delta t) = e^{H_{\mathrm{in}} \Delta t}  \sim
1/\lambda^2 \sim 2 \times 10^{26}$, for $\Delta t \sim 1/H_{\mathrm{in}} \times \ln (1/\lambda^2)$,
where $H_{\mathrm{in}}$ is the Hubble parameter as fixed by the initial vacuum energy
density of the inflaton field, which is proportional to $\lambda$.
}.

To recapitulate, the idea involves assumptions and numerical estimates. The
unusual assumption is that the $b$ field can move from a small value at a maximum of its
effective potential toward the value zero, over cosmological time intervals, but comes only
a part of the way as $t\to \infty$. 
The second essential assumption is that when the
second derivative of the effective potential with respect to the field is negative, quanta
of the $b$ field with negative squared mass are not present to induce strong long-range
forces. In the numerical estimates, an essential number is the empirically very small
self-coupling parameter $\lambda$ for the inflaton $(\phi)$ field; this parameter is common
to the $b$ field self coupling in the model.

We conclude with the remark that two general ideas seem to receive support from the
possible small decrease of $(m_p/m_e)$ with cosmological time \cite{ref1}. One
idea is that there is a small energy scale $F_b$, associated with the early
universe \cite{ref7,ref11}, in addition to the usual very high energy scales,
i.~e.~inflaton mass and radiation temperature. Effects of the $b$ field and
of the $\phi$ field are related, when depending upon the single parameter $\lambda$ \cite{ref11}.
The above smallness of $\lambda$ from the ratio of vacuum expectation values, can give
the relatively slow evolution of the field and energy density on a short time
scale for $\phi$, and possibly on a long time scale for $b$.  An essential idea in this paper is a possible relationship
of the very small parameter $\lambda$ to a cosmological time parameter $\ol{t}$. Clearly,
if $\lambda$ were to approach zero, then the residual vacuum energy density in the $b$ field
would approach zero, and the interval $\ol{t}$ would approach infinity.
The second idea is
that the small energy scale is capable of relating a very small effective cosmological
constant today\footnote{Perhaps this small energy scale limits the contributions of
zero-point energies to a homogeneous energy density (possibly time dependent).  
}
, and a small decrease of mass over a cosmological time interval.

We thank the referee for many helpful comments.

\section*{Appendix}
The idea in this paper lends itself to a speculative conception of 
a progression of universes, without end and possibly with a most
remote beginning, if any. An indefinitely long time interval in which
the $b$ field has decreased a little,
and the matter (and entropy) density has become vanishingly small, is joined
to a negative time coordinate $((\sqrt{\epsilon}F_b)\to -(\sqrt{\epsilon}F_b))$ at
which the $\phi$ field is again established at a maximum ( $|\phi|\to \simgt
(1/\lambda^2)|\sqrt{\epsilon}F_b|$, an ``initial'' condition explicitly
related \cite{ref7} to a changing $b$ field near $t=0$), causing another
``initial'', brief inflation and a ``big bang''. This is followed by another
long time of matter dilution with a diminished vacuum energy density from a 
$b$ field $\sqrt{\epsilon}F_b$.
Our motivation is that with such a concept of progression, some universes must
acquire conditions of energy densities capable of supporting structure as we know
it (with the $b$ field related to primordial CP violation, near $t=0$), even
if such conditions were not there in the remote past.

\end{document}